\documentclass[prb,floatfix,aps,twocolumn,showpacs,groupedaddress]{revtex4-1}
\usepackage[dvips]{graphics}
\usepackage[english]{babel}
\usepackage{graphicx}
\usepackage[utf8]{inputenc}
\usepackage{amssymb}
\usepackage{epstopdf}
\usepackage{bm}
\usepackage{dcolumn}
\usepackage{epsfig}
\usepackage{latexsym}
\usepackage{amsmath}
\usepackage{color}
\usepackage{array}
\usepackage{enumerate,dsfont}

\def\XXint#1#2#3{{\setbox0=\hbox{$#1{#2#3}{\int}$ }
\vcenter{\hbox{$#2#3$ }}\kern-.5\wd0}}

\definecolor{mydarkgreen}{rgb}{0,0.5,0}

\usepackage[normalem]{ulem}

\usepackage{braket}

\begin{document}
\title{Long-Range Interaction between Charge and Spin Qubits in Quantum Dots}
\author{Marcel Serina, Christoph Kloeffel, and Daniel Loss}
\affiliation{Department of Physics, University of Basel, Klingelbergstrasse 82,
CH-4056 Basel, Switzerland}
\date{\today}
\begin{abstract}
We analyze and give estimates for the long-distance coupling via floating metallic gates between different types of spin qubits in quantum dots made of different commonly used materials. In particular, we consider the hybrid, the singlet-triplet, and the spin-$1/2$ qubits, and the pairwise coupling between each type of these qubits with another hybrid qubit in GaAs, InAs, Si, and $\mathrm{Si_{0.9}Ge_{0.1}}$. We show that hybrid qubits can be capacitively coupled strongly enough to implement two-qubit gates, as long as the distance of the dots from the metallic gates is small enough. Thus, hybrid qubits are good candidates for scalable implementations of quantum computing in semiconducting nanostructures.
\end{abstract}
\pacs{03.67.Lx, 73.21.La, 73.23.Hk}
\maketitle

\section{Introduction}
One of the most promising ways to implement the concept of  quantum computation~\cite{Fey} is to use the spins of electrons  in quantum dots as qubits~\cite{DiV}.
Quantum dots fabricated in semiconductor nanostructures are used to confine electrons, which can then be manipulated and measured by electrical gates~\cite{RMP1}.
Many generalizations of qubits in quantum dots have been proposed and explored over the years~\cite{KloefRev, SiCMOS, SiScalVeldh}.
A recent interesting addition is a proposal for a so-called hybrid qubit~\cite{PRLHyb, PRL, Kim1, Kim2}  formed in a double quantum dot (DQD) by different states of three electrons, where both charge and spin degrees of freedom of the electrons play a role. Since all types of spin qubits have their specific weaknesses and strengths, it is useful to ask if one can combine various types of such qubits  to make optimal use of their particular advantages.

Furthermore, to successfully implement quantum computation, scalable architectures consisting of many qubits are needed. One of the possibilities to achieve such a goal is to couple the qubits in quantum dots over a long distance by using a metallic floating gate~\cite{PRX}. In this proposal, quantum dots could be hundreds of micrometers apart from each other, coupled in a similar way as electronic components are coupled by wires. 

Here, we extend the proposal of Ref.~\onlinecite{PRX} and study the long-distance coupling between the newly introduced hybrid qubit\cite{PRLHyb, PRL, Kim1, Kim2} and other types of qubits theoretically. For this, we consider two lateral DQDs\cite{RMP1,PettaSci, RMP2, MauneNat, KloefRev} that are separated so far from each other that tunnel coupling is impossible. A floating metallic gate enables a capacitive coupling that depends on the positions of the DQDs with respect to the gate and that can therefore be controlled electrically. We study different host materials such as Si or GaAs, analyze the dependence of the qubit-qubit interaction on several system parameters, and find that strong interactions are achievable which can be used to implement entangling two-qubit gates with short operation times.

There are three different qubit combinations we consider. First, we study the case of two hybrid (H) qubits\cite{PRLHyb, PRL, Kim1, Kim2}. In the second case, one DQD contains two and the other DQD three electrons, i.e., one DQD hosts a singlet-triplet (ST) qubit\cite{PRLST, STLongT, PettaSci} and the other one a hybrid qubit. In these two cases a spin-orbit interaction (SOI) is not required, since the qubit-qubit coupling can be induced solely by the Coulomb interaction and the Pauli exclusion principle. In the third case, we consider a system where one of the DQDs is singly occupied forming the spin-1/2 (LD) qubit~\cite{DiV, KoppensNat}  and the second one is triply occupied, i.e., a hybrid qubit. Each of these three cases will be studied separately. However, it is important to mention that experimentally it is possible to realize the different schemes within the very same experimental setup by only modifying the voltages on the gates. 

The paper is organized as follows. First in Sec.~\ref{H-H} we focus on the H-H and H-ST qubit-qubit couplings in GaAs. Then, in Sec.~\ref{H-LD} we calculate the H-LD coupling in GaAs, and in Sec.~\ref{depMatSetup} we analyze the dependence on the setup geometry and compare all the couplings in four different semiconducting materials commonly used for the fabrication of quantum dots, namely GaAs, InAs, Si, and $\mathrm{Si_{0.9}Ge_{0.1}}$. Finally, in Sec.~\ref{Conc} we summarize and give some conclusions. A detailed derivation of the electrostatic coupling via floating gates, the expansion of the electrostatic potential and details of a used Schrieffer-Wolff transformation are appended.

\section{H-H and H-ST coupling}\label{H-H}

\begin{figure}[h!]
    \renewcommand\figurename{Figure} 
\centering
\includegraphics[width=\columnwidth]{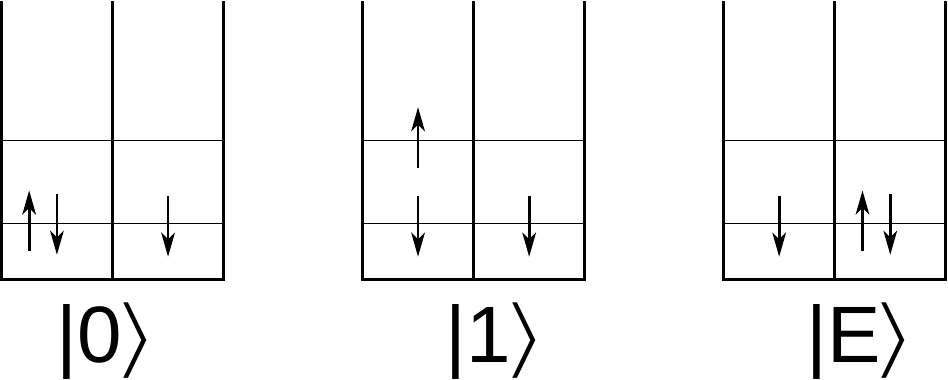}
  \caption{\label{HQ} States of the double dot hybrid qubit~\cite{PRLHyb, PRL, Kim1, Kim2}. The logical state $\ket{0}$ is defined as $\ket{0}=\ket{S}\ket{\downarrow}$, where $\ket{S}$ is the singlet of the two electrons in the ground state of the first dot and $\ket{\downarrow}$ denotes the electron with spin down in the second dot. The logical state $\ket{1}$ is $\ket{1}= \sqrt{1/3} \ket{T_0}\ket{\downarrow} - \sqrt{2/3}\ket{T_{-}}\ket{\uparrow}$, where $\ket{T_0}$ ($\ket{T_{-}}$) is the two-electron triplet state in the first dot with spin projection zero ($-1$) along the quantization axis, and $\ket{\uparrow}$ denotes the electron with spin up in the second dot. Our diagram for $\ket{1}$ is a simplified sketch of $\ket{T_0}\ket{\downarrow}$. The state $\ket{E}=\ket{\downarrow}\ket{S}$ depicted on the right is used as an intermediate state to prepare the logical states. We note that each of these three states can be written in the simple form ``orbital part $\times$ spin part'', because additional corrections are negligible in our calculations of the coupling strengths.}
\end{figure}

The DQD 
hybrid qubit, introduced recently in Ref.~\onlinecite{PRL},
operates on three different three-electron-states as illustrated in Fig.~\ref{HQ}.
In order to implement a universal set of quantum gates, one must be able to couple such qubits to each other pairwise~\cite{PRABarenco}. 

The idea pursued here is to couple two such hybrid-qubits over a (possibly long) distance via a floating gate in the way proposed in Ref.~\onlinecite{PRX} and to estimate the strength of the resulting coupling between them. This setup is promising due to its simplicity and scalability and provides a pathway for implementing large-scale quantum computing architectures. The coupling is mediated through the Coulomb interaction of the induced charges on the floating gate caused by the charge distributions in the qubits. As the different states of the hybrid qubit have different charge distributions (in contrast to qubits based on the spin states of a single quantum dot when SOI is absent), it is sufficient to take into account the Coulomb interaction in order to estimate the coupling. Thus, one may assume that the hybrid qubit is a charge qubit and may disregard the spin degrees of freedom (which are on the other hand important for the noise and decoherence description).

\begin{figure}[h!]
    \renewcommand\figurename{Figure} 
\centering
\includegraphics[width=\columnwidth]{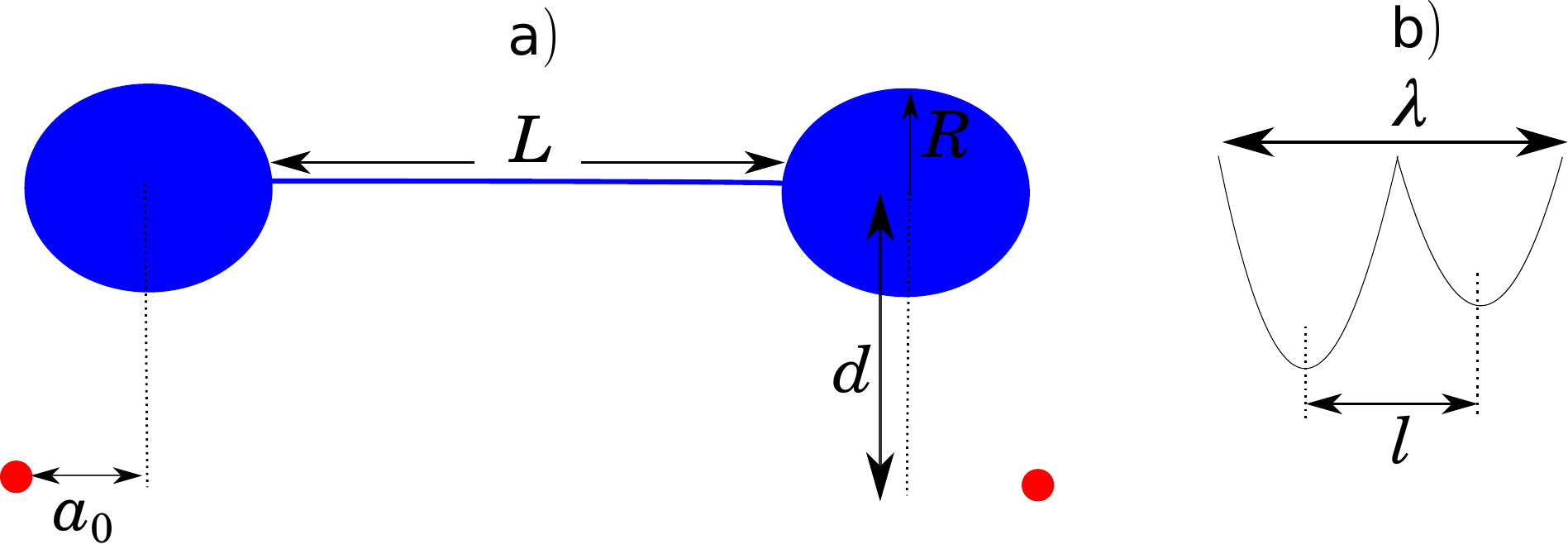}
  \caption{\label{Setup} 
   a) Setup consisting of a floating metallic gate (blue) and two DQDs (red dots). b) Sketch of the confinement potential of a DQD. Here, $a_0$ is an in-plane distance between the floating gate disc center and the DQD, $d$ is the vertical distance between the floating gate disc center and the DQD, $R$ is the radius of the floating gate disc, $L$ is the length of the section connecting the discs, $R_w$ is the radius of this section (modelled as a wire), $\lambda$ is the length of the DQD and $l$ is the distance between the DQD minima.}
\end{figure}

Following Refs.~\onlinecite{PRX,Sten2006}, we now describe the setup analytically. The induced charge in a disc of the floating gate due to an electron at position ${\bm r}$ is
\begin{equation}
Q_{\rm ind}(\bm r) = \frac{2 e}{\pi}\arcsin(R/\xi_{\bm r}),
\label{q_ind}
\end{equation}
where $e$ is the elementary positive charge and $R$ is the radius of the disc. The ellipsoidal coordinate $\xi_{\bm r}$ is given by \footnote{We note that ellipsoidal coordinates found in the literature sometimes have different underlying definitions. Details about the $\xi_{\bm{r}}$ used here are provided in Ref.~\onlinecite{Sten2006}.}  
\begin{eqnarray}
  \label{ksir} 2 \xi_{\bm r}^2 &=& R^2 + d^2 + x^2 + y^2 \\ 
  & & + \sqrt{\left(R^2 + d^2 + x^2 + y^2\right)^2 - 4 R^2 \left(x^2 + y^2\right)} . \nonumber
\end{eqnarray}
Assuming that the electron occupies a quantum dot (or DQD in our case) within the two-dimensional electron gas (2DEG) of a heterostructure, the vertical distance $d$ between the electron and the disc corresponds to the distance between the floating gate and the 2DEG, as illustrated in Fig.~\ref{Setup}a. The coordinates $x$ and $y$ describe the in-plane position of the electron, where the origin $0 = x = y$ was chosen below the center of the disc.      

When one electron is located near the first disc of the floating metallic gate and another electron is located near the second disc, the two electrons are electrostatically coupled and the associated potential energy is\footnote{Our formula differs from the one given in Ref.~\onlinecite{PRX} by a factor of 1/2. For the details see Appendix~\ref{Appendix_Details}}
\begin{equation}
v(\bm r_1,\bm r_2)=\frac{\alpha_q Q_{\rm ind}(\bm
  r_1) Q_{\rm ind}(\bm r_2) }{8 \epsilon_{r} \epsilon_{0} R} ,
  \label{Vr1r2}
\end{equation}
where $\bm{r}_1$ and $\bm{r}_2$ are the positions of the two electrons with respect to their nearby disc, $\epsilon_{0}$ is the vacuum permittivity, $\epsilon_{r}$ is an effective relative permittivity that depends on the details of the setup, and $\alpha_q = C_d/(C_w + 2C_d)$ is a charge-distribution factor that is determined by the capacitances $C_d$ and $C_w$ of the discs and the connecting wire, respectively, of the floating gate~\cite{PRX}. A detailed derivation of Eq.~(\ref{Vr1r2}) is provided in Appendix~\ref{Appendix_Details}. 

In order to obtain an estimate of the interaction we need to model the charge distributions in the DQDs for the different logical states. Electrons in the DQDs are trapped inside  the two overlapping two-dimensional (2D) parabolic potential wells (see Fig.~\ref{Setup}b), thus their orbital states in each quantum dot are in good approximation the same as the eigenstates of the 2D simple harmonic oscillator (SHO)~\cite{Burkard99}. 
The electron density in a quantum dot, which differs from the charge density only by the factor $-e$ (electron charge), can therefore be modeled by summing up the probability densities of the occupied SHO states.   

The probability density in the ground state of a SHO is Gaussian and in the first excited state~\footnote{In a 2D parabolic potential with circular symmetry, there are two degenerate orbital states that correspond to a first excited state. We assume here that this degeneracy is lifted and consider only one excited state, which is justified because asymmetries in the DQD potential can be generated electrically with the gates that control the in-plane confinement.} we have two disjoint peaks shifted with respect to the center of the well to both sides by $\lambda_0 = \sqrt{\hbar/ (m^{*}\omega)}$, where $\hbar \omega$ is the single-particle level spacing in the quantum dot and $m^{*}$ is the effective mass of the electron. We define $\lambda = l + 2 \lambda_0$ as the length of the DQD, where $l$ is the distance between the minima of the potential (see Fig.~\ref{Setup}b). Using a model as described above, with 2D continuous charge distributions in the plane of  a 2DEG, we obtain the following expression for the electrostatic coupling of the two states  $\ket{i}$, $\ket{j}$ with electron densities $\rho_{i}(\bm{r}_1)$, $\rho_{j}(\bm{r}_2)$,
\begin{eqnarray}
  \label{Vij}
  V_{ij} &=& \frac{\alpha_q }{8 \epsilon_{r} \epsilon_{0} R} \\
  & & \times \int\mathrm{d}^{2}\bm{r}_{1} \int\mathrm{d}^{2}\bm{r}_{2}\, Q_{\rm ind}(\bm{r}_{1})Q_{\rm ind}(\bm{r}_{2}) \rho_{i}(\bm{r}_{1}) \rho_{j}(\bm{r}_{2}) . \nonumber
\end{eqnarray}
These four-dimensional integrals can be evaluated numerically. However, in order to gain more insight, we will also evaluate the $V_{ij}$ analytically by substituting the Gaussian wave functions by delta functions, thus considering the interaction of point charges.

\begin{figure}[h!]
    \renewcommand\figurename{Figure} 
\centering
\includegraphics[width=\columnwidth]{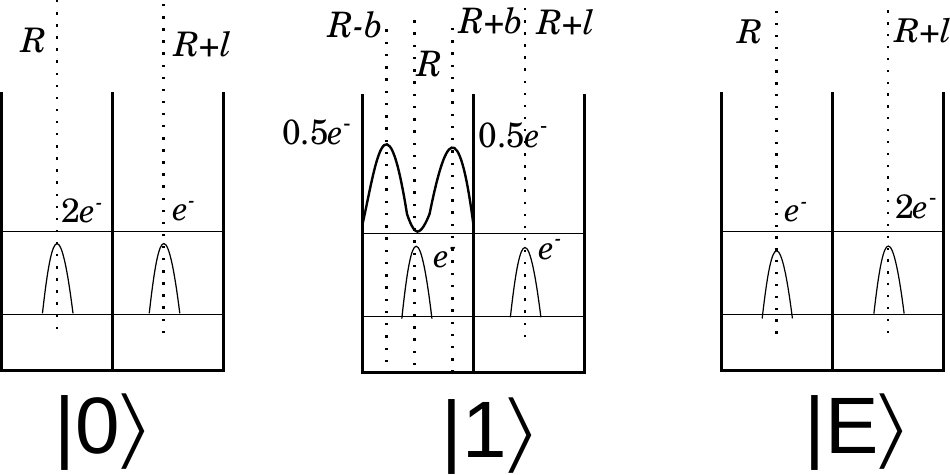}
  \caption{\label{CD} Charge distributions in the logical states of the DQD forming the hybrid qubit. In the point charge approximation we model the charge distribution via delta functions localized at the positions ($x$ coordinates) shown by dotted lines, with the corresponding charges written in multiples of the elementary charge $e^{-}=-e$. In the numerical integration we use for the charge distributions the corresponding eigenfunctions of the parabolic potential.}
\end{figure}

The electron densities in Eq.~(\ref{Vij}) depend not only on the DQD potentials but also on the positions of the DQDs with respect to the discs of the floating gate. For our estimate of the feasible coupling strengths, we assume that the two DQDs have the same dimensions and the same relative positions. More precisely, we choose the two minima of a DQD potential to lie on the $x$ axis (which can be an arbitrary in-plane direction due to the circular symmetry of the discs and which may, in fact, be different for each disc), with the first minimum at $x = a_0$ and the second minimum at $x = a_0 + l$. The parameter $a_0$ corresponds to the in-plane distance from the disc center, as sketched in Fig.~\ref{Setup}. In order to enable two-qubit gates with short operation times, this distance should be chosen such that the energies $V_{ij}$ differ strongly when changing the qubit states. In Ref.~\onlinecite{PRX}, it was shown that the floating-gate-mediated coupling between two charges is most sensitive to small variations in their positions when the charges are placed below the edges of the discs. Therefore, we set $a_0 = R$ in our model, and in our case of DQDs we choose the dots below the edges of the discs ($x=R$) to be the ones whose charge distributions depend strongest on the qubit states. Our assumptions are illustrated in Fig.~\ref{CD}, which shows both the point charge approximation and the continuous charge distribution for a hybrid qubit.

Now we are prepared to calculate the matrix elements of the electrostatic coupling
\begin{equation}
V = \sum_{m,n} v(\bm{r}_m,\bm{r}_n)
\label{eq:VSigma}
\end{equation} 
between the different states of two DQDs. In Eq.~(\ref{eq:VSigma}), the sum over $m$ and $n$ runs over all possible pairs of electrons that do not occupy the same DQD and $v$ is the electrostatic electron-electron coupling introduced in Eq.~(\ref{Vr1r2}). As the Coulomb interaction does not contain any spin operators, the spins of the states are always conserved. Thus, because the logical states are orthogonal in spin space, we get (where $i,j,k,l=0,1,E$)
\begin{eqnarray}
  \left\langle i,j\right| V \left|k,l\right\rangle =\delta_{ik}\delta_{jl}V_{kl}.
  \label{V}
\end{eqnarray}
The used notation $\left|k,l\right\rangle$ for a two-qubit state means here that the qubit at the first (second) disc is in state $\ket{k}$ ($\ket{l}$). 
As an example we can write the expression for one of the matrix elements in the 
point charge approximation as follows
\begin{eqnarray}
 \left\langle 0,1\right| V \left|0,1\right\rangle
 &=& v(R-b,R)+\frac{1}{2} v(R-b,R+l)\nonumber\\
 & & + v(R+b,R) + \frac{1}{2} v(R+b,R+l)\nonumber\\
 & & + 2 v(R,R) + v(R,R+l)\nonumber\\
 & & + 2 v(R+l,R) + v(R+l,R+l),
  \label{V01}
\end{eqnarray}
where $v(x_1,x_2)$ denotes the Coulomb interaction of two  electrons placed at the positions $x_1,x_2$. The charge in the hybrid qubit is equal to three elementary charges, thus the sum of the prefactors of all terms in Eq.~(\ref{V01}) must be nine.

Due to the aforementioned orthogonality in spin space, projection of $V$ onto the subspace spanned by the two-qubit states $\ket{0,0}$, $\ket{0,1}$, $\ket{1,0}$, and $\ket{1,1}$ yields a diagonal 4$\times$4 matrix
\begin{eqnarray}
  V = \mathrm{diag}\left(V_{00},V_{01},V_{01},V_{11}\right),
  \label{Vsbsp}
\end{eqnarray}
where we used $V_{01}=V_{10}$.
We can construct a two-qubit gate that is equivalent to the CNOT gate up to single-qubit operations as follows~\cite{PRA},
\begin{gather}
 \mathcal{H} = C \sigma^z_1\sigma^z_2 , \label{Hzz_Ham}  \\
e^{i\mathcal{H}\frac{\pi}{4C}} = e^{i\frac{\pi}{4}}\left(\begin{array}{cccc}
1 & 0 & 0 & 0\\
0 & -i & 0 & 0\\
0 & 0 & -i & 0\\
0 & 0 & 0 & 1
\end{array}\right).
  \label{Hzz}
\end{gather}
When the matrix in Eq.~(\ref{Vsbsp}) is rewritten in terms of Pauli matrices, one obtains
\begin{eqnarray}
V &=& \frac{V_{00}+V_{11}+2V_{01}}{4}+\frac{V_{00}-V_{11}}{4} \left( \sigma^z_1 + \sigma_2^z \right)  \nonumber  \\
& & + \frac{V_{00}+V_{11}-2V_{01}}{4}\sigma^z_1\sigma^z_2.  
\label{Vdecomp}
\end{eqnarray}
The last term in this decomposition enables us to implement the CNOT gate~\cite{DiV}, 
and from comparison with Eqs.~(\ref{Hzz_Ham}) and~(\ref{Hzz}) we therefore identify 
\begin{equation}
C_\textrm{H-H} = \frac{V_{00}+V_{11}-2V_{01}}{4}
\end{equation}
as the qubit-qubit coupling in the H-H setup.

We choose the parameters of the setup as follows. For the spatial dimensions of the DQDs we assume $\lambda=250\mbox{ nm}$ and $l = 150 \mbox{ nm}$, furthermore $d = 100\mbox{ nm}$ for the vertical distance and $R = a_0 = 4d$. We estimate $\alpha_q = 0.19$ from \cite{PRX}
\begin{equation}
\alpha_q \approx \frac{4 R \ln(L / R_w)}{\pi L + 8 R \ln(L / R_w)} , 
\end{equation}
with $L = 10\mbox{ $\mu$}$m and $R_w = 30\mbox{ nm}$ as the length and radius, respectively, of the thin wire that connects the two discs. Considering GaAs as the host material, we use $\epsilon_r = 13$ (the effective value may in fact be smaller, which would increase the coupling) and note that $\lambda_0 = 50\mbox{ nm}$ and the effective electron mass result in $\hbar\omega \simeq 0.5\mbox{ meV}$. With these values, which are similar to those in Ref.~\onlinecite{PRX}, we obtain $C_\textrm{H-H} = 2.0\times
10^{-10}\mbox{ eV}$ from the point charge approximation, while $C_\textrm{H-H} = 2.6\times 10^{-10}\mbox{ eV}$ from the continuous charge distributions.
 
Since these results agree very well, we conclude that the numerical integration is well reproduced by the point charge approximation. One can significantly increase the coupling by varying the setup parameters. The coupling is most significantly dependent on the vertical distance $d$. Assuming that $d=10\mbox{ nm}$ can be realized, we are able to obtain a coupling strength of $C_\textrm{H-H} = 1 \mbox{ $\mu$eV}$  (see Sec.~\ref{depMatSetup}).

Next we study the interaction between the ST qubit, i.e., singlet-triplet states of  two electrons in the DQD, and the hybrid qubit, i.e., three electrons in the DQD with states  $\ket{0}$, $\ket{1}$, and $\ket{E}$, see Fig.~\ref{CD}. We assume the DQD for the ST qubit to be in the strongly detuned regime, thus the singlet state is effectively a (2,0) charge state and the triplet state is effectively a (1,1) charge state  (excited orbital states of the dots that form the DQD cannot be occupied). Then we are able to calculate the matrix elements, for instance
\begin{eqnarray}
\left\langle S,1\right|V\left|S,1\right\rangle &=& 2 v(R,R) + v(R,R-b)  \nonumber \\
& & + v(R,R+b) + 2 v(R,R+l) .
\label{VS1}
\end{eqnarray}
Proceeding analogously to the case of two H qubits, we find a qubit-qubit interaction of the form $C_\textrm{H-ST} \sigma^z_1\sigma^z_2$, where we obtain  $C_\textrm{H-ST} = 5.5\times 10^{-9}\mbox{ eV}$ for the point
charges and $C_\textrm{H-ST} = 1.4\times 10^{-8}\mbox{ eV}$ for the continuous charge distributions. As a check of the numerics we additionally calculate the ST-ST coupling with the continuous distributions  and compare it to the expected value 
of $10^{-5}$--$10^{-6}\mbox{ eV}$ obtained from Ref.~\onlinecite{PRX}. 
We get~\cite{Note3} $C_\textrm{ST-ST} = 7.5\times 10^{-7}\mbox{ eV}$. 

It is then interesting to ask why the obtained H-H and H-ST couplings are much smaller than in the ST-ST case. When we look at the logical states $\ket{0}$ and $\ket{1}$ of the hybrid qubit, we can see that the charge distribution in the second quantum dot is the same in both logical states, thus the second quantum dot gives us no contribution to the coupling, in contrast to the states of the ST qubit. Moreover, the charge difference between the singlet and the triplet states in the single dot is small, $\delta q=4.7\times 10^{-5}e$ according to the calculations in Ref.~\onlinecite{arXiv}, in contrast to the charge difference between the singlet and triplet states in the ST qubit where $\delta q=2.7\times 10^{-2}e$. 

It is worth mentioning that when we use the state $\left|E\right\rangle$ instead of $\left|1\right\rangle$ as a computational basis state for the hybrid qubit, we obtain much larger couplings due to the different charge configurations in both wells in the different logical states. For H-H and H-ST we get about the same coupling strengths, 
$C_\textrm{H-H} \approx C_\textrm{H-ST} \approx 0.1\mbox{ $\mu$eV}$,
which is on the order of what one would expect for the charge qubit. In the next section we will pursue this idea further, because in order to implement the H-LD coupling it will be necessary to work in the $\ket{0}$, $\ket{E}$ basis. 

Finally, we note that a magnetic field is not needed to obtain the H-H, H-ST, and ST-ST couplings. However, a magnetic field is usually applied in order to energetically separate the qubit subspace from other states. The presence of an out-of-plane field, if required, can easily be included by using states of Fock-Darwin type, whose spatial extent decreases (effective confinement increases) with increasing field strength~\cite{PRB,KouwRepProg}. When the magnetic field lies in the plane of the 2DEG, which is the case in most present-day experiments~\cite{MauneNat, ShulSci, PRLHyb, STLongT, MartinsPRL} , its effects on the orbital states are negligible because of the strong and narrow vertical confinement potential. As a consequence, neither magnetic fields nor SOI are important for the estimates of this section and we can omit them in the calculations. This is in stark contrast to the LD qubits discussed in the next section, where the qubit-qubit coupling depends strongly on both the SOI and the magnetic field.

\section{H-LD coupling}\label{H-LD}

In this section we want to evaluate the coupling between the hybrid qubit in a DQD and the LD qubit in a single quantum dot. Given the coordinate system introduced in the previous section, we assume that the center of the quantum dot that hosts a LD qubit is located at $x = a_0$ and $y=0$. For the sake of simplicity in the notation, the coordinate system which we use in the present section is shifted such that its origin $0 = x = y$ and the center of the quantum dot coincide. 

The Hamiltonian of the single quantum dot reads
\begin{equation}
H_{\rm LD} = H_{0} + H_{Z}+H_{\rm SOI} 
\label{HLD}
\end{equation}
with
\begin{gather}
H_{0} = \frac{p_{x}^{2}+p_{y}^{2}}{2m^{*}} + \frac{m^{*} \omega^2}{2} \left( x^2 + y^2 \right), \\
H_{Z}=\frac{g\mu_B}{2} \bm{B}\cdot\bm{\sigma}, \\
H_{\rm SOI} = \alpha \left( p_{x} \sigma_{y} - p_{y} \sigma_{x} \right).
\end{gather}
The $p_x$ and $p_y$ are kinetic momentum operators, $g$ is the effective electron g-factor, and $\bm{\sigma}$ is the vector of Pauli matrices $\sigma_{x,y,z}$. Let us assume that the magnetic field lies in the plane of the 2DEG, i.e., $\bm{B} = \left(B_x, B_y, B_z\right) = B \left(\cos\phi, \sin\phi, 0 \right)$, where $B = |\bm{B}|$. In this case, orbital effects due to $\bm{B}$ are negligible and one can simply use $p_{x,y} = - i \hbar \partial_{x,y}$. The Zeeman energy induced by $\bm{B}$ is $E_Z = g \mu_B B$. The Rashba SOI parameter is defined as $\alpha=\frac{\hbar}{m^{*}\lambda_{\rm SO}}$, where $\lambda_{\rm SO}$ is the spin-orbit length. We assume that the SOI is relatively weak and are interested in the regime $1 \gg \frac{\left| E_Z \right|}{\hbar\omega} \gg \frac{\lambda_0}{\left| \lambda_{\rm SO} \right|}$. (We allow $E_Z$ and $\lambda_{\rm SO}$ to be positive or negative depending on the sign of $g$ and $\alpha$, respectively. As these signs, however, cannot affect the absolute values of the qubit-qubit couplings in our model, the parameters in our calculations will always be chosen positive.) In order to remove $H_{\rm SOI}$ in the first order, we can therefore employ a Schrieffer-Wolff (SW) transformation~\cite{PRB}. The details are described in Appendix~\ref{Appendix_SW}.

The floating gate couples the three electrons forming the hybrid qubit, labeled by positions $\bm{r_1,r_2,r_3}$, to the electron of the LD qubit, labeled by $\bm{r_e}$, via the potential 
\begin{equation}\label{VHybLD}
V(\bm{r_e,r_1,r_2,r_3})= v(\bm{r_e,r_1}) + v(\bm{r_e,r_2}) + v(\bm{r_e,r_3})
\end{equation}
from Eq.~(\ref{eq:VSigma}). By expanding this electrostatic potential around the positions where the confining potential of the dots is minimal, we obtain an effective term $H_{\rm int}$ that describes the floating-gate-mediated interaction between the electrons,  
\begin{eqnarray}
H_{\rm int} &=& \eta x_e(x_1+x_2+x_3) , \label{Vexpanded} \\
\eta &=&
\left. \frac{\alpha_q}{8 \epsilon_{r} \epsilon_{0} R}\left(\frac{\partial Q_{\rm ind}}{\partial x}\right)^2 \right|_{\bm r_{\rm min}} .
\label{ExpCoef}
\end{eqnarray}
The details of this expansion are provided in Appendix~\ref{Appendix_PotExpansion}. Remarkably, the term $H_{\rm int}$ is equivalent to that of an electric field applied along the DQD axis, and the strength of this electric field depends on the coordinate $x_e$ of the electron in the quantum dot for the LD qubit. We are now interested in the energy difference 
\begin{equation}
J_{01}=\bra{0}H_{\rm int}\ket{0}-\bra{1}H_{\rm int}\ket{1}
\end{equation}
which $H_{\rm int}$ causes between the logical states $\ket{0}$ and $\ket{1}$ of the hybrid qubit. As sketched in Fig.~\ref{CD}, the charge distributions of the states $\ket{0}$ and $\ket{1}$ differ only in one valley of the DQD, because one of the two electrons that occupy this valley is raised to an excited orbital state when $\ket{0}$ changes to $\ket{1}$. Consequently, we find $J_{01} = 0$ in our model, as it is known that an electric field applied to a SHO leads to a constant shift of all energy levels. It has already been proposed~\cite{PRL} to use the state $\ket{E}$ instead of $\ket{1}$ to perform a two-qubit gate as the charge difference between those states is large and one can easily pulse the state $\ket{1}$ into $\ket{E}$. One finds that
\begin{equation}
J_{0E} = \bra{0} H_{\rm int} \ket{0} - \bra{E} H_{\rm int} \ket{E}
\end{equation}
is non-zero, as a qubit based on the states $\ket{0}$ and $\ket{E}$ is effectively a charge qubit, see Fig.~\ref{CD}.

The orbital wave functions of the states $\ket{0}$ and $\ket{E}$ in the DQD are well approximated by
\begin{eqnarray}
\psi_0 &=& \psi_L(x_1, y_1)\psi_L(x_2, y_2)\psi_R(x_3, y_3), \\
\psi_E &=& \psi_L(x_1, y_1)\psi_R(x_2, y_2)\psi_R(x_3, y_3), 
\end{eqnarray}
where
\begin{eqnarray}
\psi_L(x,y) &=& \sqrt{\frac{m^*\omega}{\pi\hbar}} e^{-\frac{m^*\omega}{2\hbar}(x^2+y^2)} , \\ 
\psi_R(x,y) &=& \psi_L(x - l, y).
\end{eqnarray} 
Using these orbital wave functions one gets $J_{0E} = - \eta l x_e /2$ which corresponds to an interaction of the form
\begin{equation}
H_{\rm int}= - \frac{\eta l x_e \tau_z}{4},
\label{eq:HintTauzXe}
\end{equation}
where the Pauli operator $\tau_z$ satisfies $\tau_z \ket{0} = \ket{0}$ and $\tau_z \ket{E} = - \ket{E}$.

Finally, we can calculate the H-LD qubit-qubit coupling by considering the aforementioned SW transformation \cite{PRB}. As explained in detail in Appendix~\ref{Appendix_SW}, the antihermitian operators 
\begin{eqnarray}
S_1 &=& \frac{i}{\lambda_{\rm SO}} \left(x \sigma_y - y \sigma_x \right) ,
\label{eq:S1Maintext} \\
T_1 &=& \frac{E_Z \lambda_0^2}{i \hbar^2 \omega \lambda_{\rm SO}} \left(p_x \cos\phi + p_y \sin\phi \right) \sigma_z
\label{eq:T1Maintext}
\end{eqnarray}
are chosen such that the eigenstates of the perturbatively transformed Hamiltonian $H_{\rm LD} + [S_1 + T_1, H_{\rm LD}]$ are eigenstates of the Pauli operator $\sigma_{z^\prime} = \sigma_x \cos\phi + \sigma_y \sin\phi$, apart from small (higher-order) corrections. When we apply the same transformation to $H_{\rm int}$, keeping in mind that $x_e$ in Eq.~(\ref{eq:HintTauzXe}) corresponds to $x$ in Eqs.~(\ref{eq:S1Maintext}) and (\ref{eq:T1Maintext}), we identify
\begin{equation}
[T_1,H_{\rm int}] = C_\textrm{H-LD} \sigma_{y^\prime}\tau_z  
\end{equation}
as the H-LD coupling. We note that the form of this qubit-qubit coupling differs from those in Sec.~\ref{H-H} because the axes of the Pauli operators $\sigma_{z^\prime}$ and $\sigma_{y^\prime} = \sigma_{z}$ are perpendicular to each other. The strength of the H-LD coupling is 
\begin{equation}\label{HLDCoupl}
C_\textrm{H-LD} = \left.\frac{\alpha_q \lambda_0^2 l E_Z\cos \phi }{32 \epsilon_{r} \epsilon_{0} R\lambda_{\rm SO}\hbar\omega}\left(\frac{\partial Q_{\rm ind}}{\partial x}\right)^2 \right|_{\bm{r}_{\rm min}} ,
\end{equation}
and we recall that $\lambda_0 = \sqrt{\hbar/ (m^{*}\omega)}$.

In this calculation we included Rashba SOI only. One can perform the SW transformation including both Dresselhaus and Rashba SOIs with the following modifications. 
If the magnetic field is in the plane and if the angle $\gamma$ between the crystallographic [100] axis and the $x$ axis of our coordinate system is tuned to $\pi/4$,
it then follows from Eq.~(B10) in Ref.~\onlinecite{PRX} that the Dresselhaus SOI coefficient $\beta$ induces a simple shift of the Rashba SOI as $\tilde{\alpha}=\alpha+\beta$. The same effect can also be achieved  by choosing the in-plane magnetic field along the $x$ axis, regardless of the value of $\gamma$, and in this case we have $\tilde{\alpha}=\alpha+\beta \sin(2\gamma)$.

\section{Dependence on setup geometries and host materials}
\label{depMatSetup}

As evident from Eq.~(\ref{HLDCoupl}), we can obtain a stronger H-LD coupling by using a quantum dot material with a stronger SOI (like InAs). To verify this statement, we calculated the couplings for the fixed setup parameters mentioned before for different materials, with the following material parameters: for GaAs~\cite{PRX} $\lambda_{\rm SO} = 2.0\times 10^{-6}\mbox{ m}$, $m^{*}=0.067m_{e}$, $\hbar\omega=0.5\mbox{ meV}$, $\epsilon_{r}=13$, for InAs~\cite{RashbaInAs}
$\lambda_{\rm SO} = 1.64\times 10^{-7}\mbox{ m}$, $m^{*}=0.023m_{e}$, $\hbar\omega=1.3\mbox{ meV}$, $\epsilon_{r}=15.15$, in Si~\cite{RashbaSi} $\lambda_{\rm SO} = 2.6 \times 10^{-5}\mbox{ m}$, $m^{*}=0.26m_{e}$, $\hbar\omega=0.1\mbox{ meV}$, $\epsilon_{r}=11.7$, and for $\mathrm{Si_{0.9}Ge_{0.1}}$~\cite{RashbaSiGe}
$\lambda_{\rm SO} = 2\times 10^{-5}\mbox{ m}$, $m^{*}=0.19m_{e}$, $\hbar\omega=0.2\mbox{ meV}$, $\epsilon_{r}=12.2$. We have chosen $\omega$ in every material such that it is experimentally well accessible and keeps $\lambda_0$ close to $50\mbox{ nm}$. The Zeeman energy is always set to $E_Z=0.5\hbar\omega$. Given these parameters, the results for the qubit-qubit couplings $C_\textrm{H-H}$, $C_\textrm{H-ST}$, and $C_\textrm{H-LD}$ are listed in Table~I. However, the couplings for this setup are rather small. If we want to further enhance them, we need to consider smaller geometries. The simplest way is to equally shorten the distance $d$, the disc radius $R$, and the horizontal distance $a_0$ between the centers of disc and dot, such that $a_0 = R = 4d$ is preserved. Reducing $R$ also results in a small decrease of $\alpha_q$, which however does not dominate the parameter dependence of the qubit-qubit couplings. As expected from Eqs.~(\ref{q_ind})--(\ref{Vr1r2}), the couplings 
depend strongest on the vertical distance $d$ between the dots and the metallic gate. To quantify this dependence, we calculated the couplings again for two smaller setups. In Table~II, we present the results for currently feasible~\cite{SiCMOS} fabrication parameters with $d=57\mbox{ nm}$, $R = a_0 = 228\mbox{ nm}$, and $\alpha_q=0.13$, and in Table~III for $d=10\mbox{ nm}$, $R = a_0 = 40\mbox{ nm}$, and $\alpha_q=0.03$ to demonstrate the strong dependence on $d$. The parameters $\lambda=250\mbox{ nm}$ and $l=150\mbox{ nm}$ of the DQD and the angle $\phi=0$ of the magnetic field in the H-LD case are always fixed. 

As anticipated, a strong SOI is needed for implementing the H-LD coupling, which is why the calculated $C_\textrm{H-LD}$ in Tables~I--III are largest for InAs. In order to enhance the relatively weak H-LD couplings in Si and $\mathrm{Si_{0.9}Ge_{0.1}}$, one can add a micromagnet~\cite{MarcusMicMag, TaruchaNatPhysMicMag, TaruchaMicMag} to the setup, because the magnetic field gradient caused by a micromagnet can result in a very strong effective SOI. Considering the transformation discussed in Sec.~\ref{H-LD} and Appendix~\ref{Appendix_SW}, one obtains~\cite{AleinerSpinRel, KhaetSpinRel, FabianSpinRel}
\begin{equation}
\left[S_1, \frac{g \mu_B}{2} \bm{B} \cdot \bm{\sigma} \right] = \frac{g \mu_B B}{\lambda_{\rm SO}} \left( x \cos\phi + y \sin\phi \right) \sigma_z , 
\end{equation} 
where $S_1$ is the antihermitian operator of Eq.~(\ref{eq:S1Maintext}) and $\bm{B} \cdot \bm{\sigma} = B \left( \sigma_x \cos\phi + \sigma_y \sin \phi \right)$ due to the in-plane magnetic field. In particular, when we choose $\phi=0$ as in our calculations for $C_\textrm{H-LD}$ in Tables~I--III, the transformed Zeeman term is simply $g \mu_B B x \sigma_z/\lambda_{\rm SO}$, which corresponds to the effect of an out-of-plane magnetic field that increases linearly with the coordinate $x$. Assuming that the micromagnet provides such a magnetic field $b_{\rm MM} x$ along the $z$ axis, with $b_{\rm MM}$ as the gradient, the total (effective) gradient $b_{\rm tot}$ is therefore \cite{TaruchaMicMag}
\begin{equation}
b_{\mathrm{tot}} = b_\mathrm{MM} + \frac{2 B}{\lambda_{\rm SO}} .
\end{equation} 
Consequently, even if a material does not feature Rashba or Dresselhaus SOI at all, a spin-orbit length of the order of $2 B / b_\mathrm{MM}$ can be induced by a micromagnet (the estimate presented here applies when $|\lambda_{\rm SO}| \gg \lambda_0$). Experimental setups with $b_{\mathrm{MM}} \sim 1\mbox{ T/$\mu$m}$ are already well feasible~\cite{TaruchaMicMag}, which results in an effective spin-orbit length $\lambda_{\rm SO} \sim 2 \times 10^{-6}\mbox{ m}$ when we keep our other parameters for the Si and $\mathrm{Si_{0.9}Ge_{0.1}}$ setups unchanged. In both Si and $\mathrm{Si_{0.9}Ge_{0.1}}$, we find that the considered gradient from the micromagnet increases the H-LD couplings by an order of magnitude. For instance, using the setup with $d = \mbox{10 nm}$ and micromagnets to enhance the SOI, one can reach $C_\textrm{H-LD} \sim 0.5 \mbox{ $\mu$eV}$ for Si 2DEGs.

For a comparison of the H-H and H-ST couplings in Tables~I--III, it is important to keep in mind that the materials differ from each other only in the relative permittivities $\epsilon_r$ and  effective masses $m^{*}$ in the context of this calculation. As we chose different values of $\omega$ for the different materials, such that the confinement length $\lambda_0$ is approximately constant, we find that the listed values of $C_\textrm{H-H}$ and $C_\textrm{H-ST}$ are quite similar, because it is evident from Sec.~\ref{H-H} that the couplings $C_\textrm{H-H}$ and $C_\textrm{H-ST}$ are independent of $m^{*}$ when $\lambda_0$ is fixed. Thus, the small differences that remain in the H-H and H-ST cases between the couplings obtained for the various 2DEGs are solely caused by the small variations in $\lambda_0$ and the relative permittivity.

In our analysis of the charge distributions in the states of single and double quantum dots, we assumed that all wave functions of excited states have an excited orbital part. In Si, we therefore assume that the orbital level spacing $\hbar\omega = 0.1\mbox{ meV}$ is smaller than the valley splitting, which is reasonable because reported valley splittings are of the order of \mbox{0.1--1 meV}~\cite{SiRevModPhys, SiValSplErik, SiValSplJiang, Morello}. For instance, Ref.~\onlinecite{Morello} reports an electrically tunable valley splitting in the range of \mbox{0.3--0.8 meV} for a quantum dot in Si/SiO$_2$. Consequently, forming Si quantum dots with sufficiently large valley splittings is clearly possible (analogous for $\mathrm{Si_{0.9}Ge_{0.1}}$).

At the end of this section, we would like to summarize some important dependences of the qubit-qubit couplings on the sample parameters. For the H-H coupling, we can conclude from Eqs.~(\ref{q_ind})--(\ref{Vr1r2}) that $C_\textrm{H-H} \sim \frac{1}{d^2}, \frac{1}{R}$, and the same applies of course to the H-ST coupling $C_\textrm{H-ST}$. Moreover, Eq.~(\ref{HLDCoupl}) reveals that $C_\textrm{H-LD} \sim B,  \alpha,\omega^{-2}$ for the H-LD coupling.

Finally, we wish to emphasize that additional electric gates, which are used to control and manipulate the qubits in an experimental setup, were not yet included in our calculations. As shown in Ref.~\onlinecite{PRX}, the influence of these surrounding gates can increase the qubit-qubit couplings by another two orders of magnitude, which makes the hybrid qubit a highly promising platform for quantum computation.

\onecolumngrid
\begin{center}
    \begin{table}[!htb]

\begin{tabular*}{\columnwidth}{@{\extracolsep{\fill}} l l l l l }
\hline
\hline
\rule{0pt}{2.4ex}
Coupling $C$ [$\mu$eV] & InAs & GaAs & $\mathrm{ Si_{0.9}Ge_{0.1}}$ & Si \\
\hline 
    H-H\rule{0pt}{3ex}   & $2.8\times 10^{-4}$ & $2.6\times 10^{-4}$ & $2.2\times 10^{-4}$ & $4.7\times 10^{-4}$ \\ 
    H-ST & $1.3\times 10^{-2}$ & $1.4\times 10^{-2}$ & $1.3\times 10^{-2}$ & $1.9\times 10^{-2}$ \\ 
    H-LD & $3.5\times 10^{-2}$ & $3.0\times 10^{-3}$ & $2.8\times 10^{-4}$ & $3.3\times 10^{-4}$ \\
\hline
\hline
\end{tabular*}
\caption{
Calculated strengths of the coupling between the three different types of qubits [spin-$1/2$ (LD), singlet-triplet (ST), and hybrid (H) qubit] in DQDs for $d=100\mbox{ nm}$, $\alpha_q=0.19$,  $R = a_0 = 400\mbox{ nm}$ (see Fig.~\ref{Setup}) for four different semiconducting materials commonly used for the realization of DQDs. We note that in the H-LD case, the two-qubit gate is assumed to be performed in the $\ket{0}, \ket{E}$ basis of the hybrid qubit. In all these calculations we used the continuous charge densities.} 
\label{c100}
\end{table}
\begin{table}[!htb]

\begin{tabular*}{\columnwidth}{@{\extracolsep{\fill}} l l l l l}
\hline
\hline
\rule{0pt}{2.4ex}
Coupling $C$ [$\mu$eV] & InAs & GaAs & $\mathrm{ Si_{0.9}Ge_{0.1}}$ & Si \\
\hline
H-H\rule{0pt}{3ex} & $2.7\times 10^{-3}$ & $2.6\times 10^{-3}$ & $2.2\times 10^{-3}$ & $4.5\times 10^{-3}$ \\ 
    H-ST & $6.5\times 10^{-2}$ & $7.0\times 10^{-2}$ & $6.5\times 10^{-2}$ & $9.5\times 10^{-2}$ \\ 
    H-LD & $1.3\times 10^{-1}$ & $1.1\times 10^{-2}$ & $1.0\times 10^{-3}$ & $1.2\times 10^{-3}$ \\
\hline
\hline
\end{tabular*}
\caption{
Same as in Table I for $d=57\mbox{ nm}$, $\alpha_q=0.13$ and $R=a_0=228\mbox{ nm}$.
\label{c57}
}
\end{table}
\begin{table}[!htb]
\begin{tabular*}{\columnwidth}{@{\extracolsep{\fill}} l l l l l}
\hline
\hline
\rule{0pt}{2.4ex}
Coupling $C$ [$\mu$eV] & InAs & GaAs & $\mathrm{ Si_{0.9}Ge_{0.1}}$ & Si \\
\hline 
H-H\rule{0pt}{3ex} & $3.3\times 10^{-1}$ & $3.7\times 10^{-1}$ & $3.7\times 10^{-1}$ & $4.4\times 10^{-1}$ \\ 
    H-ST & $1.3$ & $1.5$ & $1.6$ & $1.5$ \\ 
    H-LD & $5.6$ & $4.7\times 10^{-1}$ & $4.5\times 10^{-2}$ & $5.2\times 10^{-2}$ \\
\hline
\hline
\end{tabular*}
\caption{
Same as in Table I for $d=10 \mbox{ nm}$, $\alpha_q=0.03$ and $R=a_0=40 \mbox{ nm}$. In this setup, as $\lambda=250\mbox{ nm}$ the point charge approximation would be very imprecise and should not be expected to be comparable with our results obtained with the continuous densities.}
\label{c10}
\end{table}
\end{center}
\twocolumngrid

\section{Conclusions}\label{Conc}

We have studied an experimentally realizable setup which allows 
different
types of qubits in DQDs to be coupled over long distances. In particular, we have shown that using a metallic floating gate, it is possible to capacitively couple the hybrid qubit to a single spin 1/2, to a singlet-triplet qubit, or to another hybrid qubit. First we employed a simple approximation, where the charge density within the quantum dot is treated as a point charge distribution, and calculated the couplings between  two hybrid qubits as well as between the ST and the hybrid qubit. Furthermore, we have verified the validity of the point charge approximation by numerically integrating over the continuous charge distribution within each quantum dot. Next, 
we have calculated perturbatively the coupling between the LD and the hybrid qubit, assuming that the latter is based on the states $\ket{0}$ and $\ket{E}$. In order to investigate the influence of the quantum dot material and the setup geometry, we have calculated
the couplings for the four semiconducting materials GaAs, InAs, Si, $\mathrm{Si_{0.9}Ge_{0.1}}$, and the three different setups of Tables I--III.

As anticipated, the strongest H-LD coupling is obtained with InAs, which features the strongest SOI. Nevertheless, as explained in Sec.~\ref{depMatSetup}, micromagnets enable a strong, effective SOI (and therefore strong H-LD couplings) even in materials where the intrinsic SOI is weak, such as Si or $\mathrm{Si_{0.9}Ge_{0.1}}$.

For setups with a small distance $d=10\mbox{ nm}$ between the 2DEG and the floating gate, our calculated H-LD, H-ST, and H-H couplings are of the order of one microelectronvolt (see Table~III), which leads to very fast two-qubit gates with subnanosecond operation times. In fact, such short time scales may already be possible with present-day samples (see Table~II, $d = 57\mbox{ nm}$), assuming that the qubit-qubit couplings increase by two orders of magnitude, as simulated in Ref.~\onlinecite{PRX}, when all elements of the sample, especially all the electric gates, are included in the calculation.

In conclusion, we showed that hybrid qubits can be very strongly coupled over long distances via floating metallic gates, enabling long-range two-qubit gates with short operation times and all-electrical control. This applies to all of the studied materials, including Si and $\mathrm{Si_{0.9}Ge_{0.1}}$. We note that Ge and Si can be grown nuclear-spin-free and are therefore highly useful for implementing spin qubits with long dephasing times~\cite{MauneNat}. Our results prove that hybrid qubits are a promising platform for quantum information processing and quantum computation with spins in quantum dots. 

\begin{acknowledgments} 
We are especially grateful to M. A. Eriksson for interesting us in this problem and for useful comments. This work was supported in part by the Swiss NF, NCCR QSIT, SiSPIN, and IARPA. 
\end{acknowledgments}

\appendix

\section{Derivation of the electron-electron coupling mediated by the floating gate}
\label{Appendix_Details}

In this appendix we explain Eqs.~(\ref{q_ind})--(\ref{Vr1r2}) of the main text in more detail and provide an intuitive picture for the derivation of the electrostatic electron-electron interaction described by Eq.~(\ref{Vr1r2}).

We consider a thin conducting disc of radius $R$ and assume for now that it is grounded. When an electron is located nearby, the charge \cite{Sten2006} 
\begin{equation}
Q_{\rm ind}(\bm r) = \frac{2 e}{\pi} \arcsin(R/\xi_{\bm r})
\label{q_ind_App}
\end{equation}
is induced in the disc, where $e$ is the elementary positive charge. Defining the center of the disc as the origin of the coordinate system, the ellipsoidal coordinate $\xi_{\bm r} > R$ is related to the position $\bm{r} = (x,y,z)$ of the electron via
\begin{eqnarray}
\label{ksir_App} 
2\xi_{\bm r}^2 &=& R^2 + x^2 + y^2 + z^2  \\ 
& & + \sqrt{\left(R^2 + x^2 + y^2 + z^2\right)^2 - 4R^2 (x^2 + y^2)} ,  \nonumber
\end{eqnarray}
which corresponds to the solution of
\begin{equation}
\frac{x^2 + y^2}{\xi_{\bm r}^2} + \frac{z^2}{\xi_{\bm r}^2 - R^2} = 1
\end{equation}
that satisfies $\xi_{\bm r} > R$~\cite{Note2}. The Cartesian coordinate system was chosen here such that the disc lies in the $x$-$y$ plane. 
It is important to note that Eq.~(\ref{q_ind_App}) for the induced charge does not change when we allow the system to be immersed in a dielectric medium with relative permittivity $\epsilon_r$. This might initially be surprising, but it becomes easily comprehensible when one keeps in mind that the electrostatic potential due to the electron and the induced charge $Q_{\rm ind}$ must compensate at the grounded disc. Keeping $Q_{\rm ind}$ fixed, the potential of both the electron and the disc are inversely proportional to $\epsilon_r$ (which is also seen, for instance, in the capacitance $8 \epsilon_r \epsilon_0 R$ of a disc in a dielectric), and so the solution for $Q_{\rm ind}$ must indeed be independent of the relative permittivity. In our setup (Fig.~\ref{Setup}), where vacuum above the floating gates and different layers of materials may be involved, small corrections to Eq.~(\ref{q_ind_App}) can be expected which, however, will only have minor quantitative effects on our results. A detailed derivation of 
Eq.~(\ref{q_ind_App}) is provided in Ref.~\onlinecite{Sten2006}, using the aforementioned assumption that the disc is grounded. In fact, even though the attached wire and the second disc of our floating gate correspond to a finite reservoir, the assumption of a grounded disc is not satisfied because the floating gate is isolated and so the total charge in the gate must be conserved. Nevertheless, as explained in the following, Eq.~(\ref{q_ind_App}) can be exploited to calculate the steady-state charge distribution of the floating gate.

Given our gate geometry of Fig.~\ref{Setup}, we now assume that an electron is brought close to one of the discs, referred to as disc~1. After a very short (subpicosecond) time \cite{PRX}, much shorter than the typical duration of a quantum operation on a qubit,
the charge distribution in the metallic gate will have reached a steady state which can be described by the following simple requirements,
\begin{gather}
\frac{Q_{d1}}{C_d} - \frac{Q_{\rm ind,1}}{C_d} = \frac{Q_{w}}{C_w} = \frac{Q_{d2}}{C_d} ,  \label{samePot_App} \\
Q_{d1} + Q_{w} + Q_{d2} = 0 . \label{chargeCons_App}
\end{gather}
With Eq.~(\ref{samePot_App}), we exploit that the floating gate is conducting, and so the potential at disc~1, the connecting wire, and disc 2 must be the same. Equation~(\ref{chargeCons_App}) arises from charge conservation in the floating gate. The introduced quantities $Q_{d1}$, $Q_{w}$, and $Q_{d2}$ are the charges in disc~1, the wire, and disc~2, respectively. The capacitance of the wire is $C_w$ and, assuming a symmetric gate geometry, each of the two discs has the capacitance $C_d$. The term $- Q_{\rm ind,1}/C_d$ on the left-hand side of Eq.~(\ref{samePot_App}) accounts for the additional potential at disc~1 which results from the external electron. If $Q_{d1} = Q_{\rm ind,1}$, this term is exactly compensated, and so $Q_{\rm ind,1}$ can be interpreted as the charge that would be induced in disc~1 and that would remain there if the disc were grounded [see also Eq.~(\ref{q_ind_App})]. Solving the system of three equations contained in Eqs.~(\ref{samePot_App}) and (\ref{chargeCons_App}) yields the 
charge distribution in the steady state,
\begin{eqnarray}
Q_{d1} &=& Q_{\rm ind,1} \frac{C_w + C_d}{C_w + 2 C_d} , \\ 
Q_{w} &=& - Q_{\rm ind,1} \frac{C_w}{C_w + 2 C_d} , \\ 
Q_{d2} &=& - Q_{\rm ind,1} \frac{C_d}{C_w + 2 C_d} . \label{ChargeInD2_App}
\end{eqnarray}
In particular, defining
\begin{equation}
\alpha_q = \frac{C_d}{C_w + 2 C_d} , \label{AlphaQ_App} 
\end{equation} 
we see from Eq.~(\ref{ChargeInD2_App}) that the electron near disc~1 results in a charge $Q_{d2} = - \alpha_q Q_{\rm ind,1}$ in disc 2.

Next, we analyze the electric potential around the second disc, which is charged by $Q_{d2}$. When a thin conducting disc with radius $R$ and charge $Q$ is surrounded by a dielectric with relative permittivity $\epsilon_r$, the potential at a position $\bm{r} = (x,y,z)$ with $\xi_{\bm r} > R$ is 
\begin{equation}
\Phi(\bm{r}) = \frac{Q}{4 \pi \epsilon_r \epsilon_0 R} \arcsin(R/\xi_{\bm r}) .
\label{PotAroundDisc_App}
\end{equation}
Again, the ellipsoidal coordinate $\xi_{\bm r}$ is given by Eq.~(\ref{ksir_App}) if the disc lies in the $x$-$y$ plane of the chosen Cartesian coordinate system and if the center of the disc and the origin coincide. The result in Eq.~(\ref{PotAroundDisc_App}) is identical with those provided, e.g., in Refs.~\onlinecite{Sten2006, LandLif}, keeping in mind some properties of inverse trigonometric functions such as
\begin{equation}
\arcsin(R/\xi_{\bm r}) = \arctan\Bigl(R/\sqrt{\xi_{\bm r}^2 - R^2} \Bigr)
\end{equation}
and
\begin{equation}
\arcsin(R/\xi_{\bm r}) = R \int_{\xi_{\bm r}}^{\infty} \frac{d \xi}{\xi \sqrt{\xi^2 - R^2 }}
\end{equation}
for $\xi_{\bm r} > R$. In our setup, the potential within the material below disc~2 is therefore well described by
\begin{equation}
\Phi(\bm{r}_2) = \frac{Q_{d2}}{4 \pi \epsilon_r \epsilon_0 R} \arcsin(R/\xi_{\bm{r}_2}) , 
\label{PotBelowDisc2_App}
\end{equation}
where the subscript added to $\bm{r}_2$ and $\xi_{\bm{r}_2}$ refers to the second disc and where $\epsilon_r$ may be replaced by an effective value if, e.g., the floating gate is located at the sample surface (partially in vacuum) or if the sample consists of multiple layers.

Finally, we can combine the previous equations and calculate the electron-electron coupling mediated by the floating gate. Given our result for $\Phi(\bm{r}_2)$, Eq.~(\ref{PotBelowDisc2_App}), the associated potential energy $v$ of an electron below disc~2 is simply $- e \Phi(\bm{r}_2)$. Referring to the position of the electron below disc~1 as $\bm{r}_1$, the combination of Eqs.~(\ref{q_ind_App}), (\ref{ChargeInD2_App}), (\ref{AlphaQ_App}), and (\ref{PotBelowDisc2_App}) yields
\begin{eqnarray}
v(\bm{r}_{1},\bm{r}_{2}) &=& \frac{e \alpha_q Q_{\rm ind}(\bm{r}_{1})}{4 \pi \epsilon_r \epsilon_0 R} \arcsin(R/\xi_{\bm{r}_2})  \nonumber \\
&=& \frac{\alpha_q}{8 \epsilon_r \epsilon_0 R} Q_{\rm ind}(\bm{r}_{1}) Q_{\rm ind}(\bm{r}_{2}) ,
\label{Vr1r2_App}
\end{eqnarray}
which is the result shown in Eq.~(\ref{Vr1r2}) of the main text. We note that the same result is obtained when one starts the derivation at disc~2 instead of disc~1, which is also evident from the symmetric form of the final expression. As a last remark, we mention that this coupling may be interpreted as
\begin{equation}
v(\bm{r}_{1},\bm{r}_{2}) \approx \frac{Q_{\rm ind}(\bm{r}_{1}) Q_{\rm ind}(\bm{r}_{2})}{C_w + 2 C_d} 
\end{equation}
with the approximation $C_d \approx 8 \epsilon_r \epsilon_0 R$, which simply corresponds to the product of the induced charges [Eq.~(\ref{q_ind_App})] divided by the total capacitance of the floating gate.

\section{Schrieffer-Wolff transformation}
\label{Appendix_SW}

In this appendix we provide the details of the SW transformation that was performed in Sec.~\ref{H-LD} of the main text. We note that this transformation is often exploited in the literature~\cite{PRX, PRB, AleinerSpinRel, KhaetSpinRel, FabianSpinRel}.

Given the Hamiltonian in Eq.~(\ref{HLD}) and the regime $1 \gg \frac{\left| E_Z \right|}{\hbar\omega} \gg \frac{\lambda_0}{\left| \lambda_{\rm SO} \right|}$, we consider two consecutive unitary transformations in order to remove the SOI perturbatively, 
\begin{gather}
H_{\mathrm{LD}}^\prime = e^T e^S H_{\mathrm{LD}} e^{-S} e^{-T} , \\ 
S= S_1 + \ldots , \\ 
T= T_1 + \ldots ,
\end{gather}
where ``\ldots '' stands for additional terms of higher order. We choose the form of $S_1$ such that the strongest term of the SOI is eliminated,
\begin{gather}
[S_1,H_0]=-H_{\rm SOI}, \\
S_1=\frac{i}{\lambda_{\rm SO}}(x\sigma_y-y\sigma_x) .
\end{gather}
Since
\begin{eqnarray}
\nonumber [S_1,H_Z]= H_{\rm SOI}^Z = \frac{E_Z}{\lambda_{\rm SO}} \left( x \cos\phi + y \sin\phi \right) \sigma_z ,
\end{eqnarray}
we have to apply the second transformation such that $H_{\rm SOI}^Z$ is removed,
\begin{gather}
[T_1,H_0] = -H_{\rm SOI}^Z , \\
T_1 = \frac{E_Z \lambda_0^2}{i \hbar^2 \omega \lambda_{\rm SO}} \left( p_x \cos\phi + p_y \sin\phi \right) \sigma_z .
\end{gather}
Moreover, as the eigenstates of $H_{\mathrm{LD}}^\prime$ correspond essentially to the eigenstates of $H_0+H_Z$ and because we consider an in-plane magnetic field $\bm{B} = B \left(\bm{e}_x \cos\phi + \bm{e}_y \sin\phi \right)$, with $\bm{e}_i$ as unit vectors for the respective directions, we introduce new basis vectors $\bm{e}_{z^\prime} = \bm{e}_x \cos\phi + \bm{e}_y \sin\phi$, $\bm{e}_{y^\prime} =\bm{e}_z$, and $\bm{e}_{x^\prime} = \bm{e}_y \cos\phi - \bm{e}_x \sin\phi $. Consequently,
\begin{gather}
\sigma_x = \sigma_{z^\prime} \cos\phi - \sigma_{x^\prime} \sin\phi , \\
\sigma_y = \sigma_{x^\prime} \cos\phi + \sigma_{z^\prime} \sin\phi , \\
\sigma_z = \sigma_{y^\prime} , 
\end{gather}
and
\begin{equation}
H_Z = \frac{E_Z}{2} \sigma_{z^\prime} .
\end{equation}

\section{Expansion of the interaction potential}
\label{Appendix_PotExpansion}

When we expand the electrostatic potential in Sec.~\ref{H-LD} around the positions of the quantum dots, we follow the approach discussed in Appendix~B of Ref.~\onlinecite{PRX}. In the following, we focus on the expansion of the term $v(\bm{r_e},\bm{r_1})$, as the other terms $v(\bm{r_e},\bm{r_2})$ and $v(\bm{r_e},\bm{r_3})$ are expanded analogously. 

We assume that the quantum dot that hosts the LD qubit is located below the edge of one of the metallic discs. Referring to the coordinate system introduced in Sec.~\ref{H-H} (2DEG in $x$-$y$ plane, origin $0 = x = y$ below the center of the nearby disc), we thus assume that the confining potential of the quantum dot is minimal at the coordinates $x = a_0 = R$ and $y = 0$. Analogously, we consider the DQD with the hybrid qubit to be located below the edge of the other metallic disc. In our model, the coordinates of the dot that contains two electrons when the hybrid qubit is in the state $\ket{0}$ are $x = a_0 = R$ and $y = 0$. The coordinates of the second dot of the DQD are $x = a_0 + l$ and $y = 0$. Thus, keeping in mind that the origins of the coordinate systems for the LD and the hybrid qubit are related to the respective metallic discs, we find that a minimum of the confining potential occurs at the position $\bm{r}_{\rm min} = (a_0, 0) = (R, 0)$ for both qubits. 

As we are interested in the qubit-qubit coupling that results from $v(\bm{r_e},\bm{r_1})$, we are looking for terms of type $x_e x_1$, $x_e y_1$, $y_e x_1$, or $y_e y_1$ in the expansion. All the other terms up to the second power in coordinates cannot lead to a qubit-qubit interaction. Moreover, as the quantum dots are displaced by $a_0$ along the $x$ axis only, we find 
\begin{equation}
\left.\left(\frac{\partial Q_{\rm ind}}{\partial y}\right)\right|_{y=0}=0 .
\end{equation}
Thus, the expansion up to the second power in coordinates yields
\begin{equation}
v(\bm{r_e},\bm{r_1}) \approx \ldots + \eta x_e x_1,
\end{equation}
where ``\ldots '' stands for all the constant or frequency-rescaling terms that do not affect the qubit-qubit coupling and 
\begin{equation}
\eta = \frac{\alpha_q}{8 \epsilon_{r} \epsilon_{0} R} \left.\left(\frac{\partial Q_{\rm ind}(\bm{r_e})}{\partial x_e}\right)\right|_{\bm r_{\rm min}}\left.\left(\frac{\partial Q_{\rm ind}(\bm{r_1})}{\partial x_1}\right)\right|_{\bm r_{\rm min}} . \hspace{0.3cm}
\end{equation}
Thus, due to the assumed symmetry in the setup, the constant $\eta$ can be written in the short form
\begin{equation}
\eta = \frac{\alpha_q}{8 \epsilon_{r} \epsilon_{0} R} \left.\left(\frac{\partial Q_{\rm ind}}{\partial x}\right)^2 \right|_{\bm r_{\rm min}} .
\end{equation}
After such an expansion of all three terms $v(\bm{r_e},\bm{r_1})$, $v(\bm{r_e},\bm{r_2})$, and $v(\bm{r_e},\bm{r_3})$ of the interaction potential, we obtain Eqs.~(\ref{Vexpanded}) and (\ref{ExpCoef}) of the main text.

\bibliography{referencies.bib}
\end{document}